\documentclass{PoS}

\newcommand\jhep{{\sl J.\ H.\ E.\ P.\/}\ }
\newcommand\np{{\sl Nucl.\ Phys.\/}\ }

\newcommand\pr{{\sl Phys.\ Rev.\/}\ }

\newcommand\ptep{{\sl Prog.\ Theor.\ Exp.\ Phys.\/}\ }
\newcommand{\tc}{$T_c$}
\newcommand{\prenT}{P_{\rm ren}(T, t)}
\newcommand{\pren}{P_{\rm ren}(T)}
\newcommand{\gms}{g^2_{\overline{\scriptscriptstyle MS}}}
\newcommand\beq{\begin{equation}}
\newcommand\eeq[1]{\label{eq.#1}\end{equation}}
\newcommand\eq[1]{Eq.\ (\ref{eq.#1})}
\newcommand{\tav}[1]{\langle  #1 \rangle_{\scriptscriptstyle T}}
\newcommand{\tT}{\sqrt{t}T}
\newcommand{\Ebar}{\langle \underline{G_{\scriptstyle E}^2} \rangle_T}

\newcommand{\Eav}{\langle E(T, t) \rangle}
\newcommand{\Mav}{\langle M(T, t) \rangle}

\title{Looking at the deconfinement transition using Wilson flow}

\ShortTitle{Wilson flow and deconfinement}

\author{\speaker{Saumen Datta} and Sourendu Gupta\\
        Department of Theoretical Physics, Tata Institute of Fundamental  Research, Mumbai 400005, India. \\
        E-mail: \email{saumen@theory.tifr.res.in}, \email{sgupta@theory.tifr.res.in}}

\author{Andrew Lytle \\
        SUPA School of Physics and Astronomy, University of Glasgow, Glasgow G12 8QQ, United Kingdom. \\
        E-mail: \email{andrew.lytle@glasgow.ac.uk}}

\abstract{Wilson flow is an effective tool for constructing renormalized
composite operators. We explore use of the Wilson flow to construct 
renormalized order parameters for the deconfinement transition in SU(3) 
gauge theory. We discuss renormalization of the Polyakov loop, and of 
gluon condensates.}

\FullConference{34th annual International Symposium on Lattice Field Theory\\
		24-30 July 2016\\
		University of Southampton, UK}

\begin{document}

\section{Introduction}

Wilson flow, the evolution of the gauge links along the gradient of the 
gauge action, is a powerful new technique in the study of non-Abelian 
gauge theories on a lattice \cite{main,narayanan}. Some of its most common 
uses have been in scale setting \cite{main,scale} and in renormalization 
of composite operators, e.g., Refs. \cite{luscher,suzuki}. The renormalization
of the composite operators can be particularly useful in the context of finite 
temperature studies of QCD. An early example has been the calculation of
pure gauge theory thermodynamics with energy-momentum tensor renormalized using
Wilson flow \cite{flowqcd}. Here we report on the use of flow to construct
an order parameter for the pure gauge theory transition, and to study the 
behavior of gluon condensates across the transition. More details of our 
study can be found in Ref. \cite{prd}.

After Wilson flow to time $t$, the link operators are 
smeared to a size $\propto \sqrt{t}$. L\"uscher has suggested defining the
scale through a construction involving the gluon condensate. A dimensionful
quantity $t_c$ is defined as the flow time $t$ such that $\mathcal{E}(t) \ = \ 
t^2 \langle {\bf E}(t) \rangle \ = \ c$, where $c$ is a suitable number, 
\beq
{\bf E}(t) = - \frac{1}{2 L^3 T} \sum_{x, t} {\rm Tr} \ G_{\mu \nu}(x, t) 
G_{\mu \nu}(x, t),
\eeq{cond0}
$G_{\mu \nu}$ is the (discretized) field strength tensor and $L^3 T$ is the 
space-time volume. In perturbation theory \cite{luscher}
\beq
\mathcal{E}(t) = \frac{16 \pi^2}{3} \gms (\mu=1/\sqrt{8 t}) \ \left[ 1 + 0.08736 \; \gms + \mathcal{O} (g^4) \right].
\eeq{coupling}

In order that effects of the ultraviolet scale $1/a$ is suppressed, $c$
should be such that $\sqrt{t} \gg a$. For scale setting, L\"uscher has
suggested $c$=0.3; and the corresponding $t_c$ is commonly referred to as $t_0$.
At finite temperature, $T$ sets the energy scale of interest;
since flow strongly suppresses energy scales $> 1/\sqrt{t}$, ideally
for thermal physics one would like to have
\beq
T \ll \frac{1}{\sqrt{t}} \ll \frac{1}{a}. 
\eeq{range}
In typical finite temperature lattice studies at present, 
$a \gtrsim 1/16 T$; so the strong inequalities in \eq{range}
can at most mean ``smaller by a factor $\sim$ 4''.
If we want to keep c fixed while the temperature is 
changed, \eq{range} can be satisfied 
only if $1/(N_t \sqrt{t_c}) \ll T \ll 1/\sqrt{t_c}$ for all $T$ and
$N_t$. In \cite{flowqcd}, $\sqrt{t}T = b$ was 
fixed as one changes $T$. An exploration of these strategies 
will be reported as part of the study.

We study flow on finite temperature lattices in the 
temperature range between 0.9 \tc\ and 3.1 \tc\ employing four sets of
lattices, corresponding to $N_t=$ 6, 8, 10 and 12. In each set, temperature 
is changed by changing $\beta$; $T_c$ is set from the peak of the Polyakov 
loop susceptibility and the relative temperature $T/T_c$ in the other lattices
is set using Wilson flow. The $N_t=12$ 
lattices are new; details of the other sets can be found in \cite{prd}. 
First, we discuss the Polyakov loop,
which is the order parameter for the deconfinement transition, but is
highly singular as one takes the continuum limit. We discuss in the
next section the use of flow to construct a continuum order
parameter, referred to here as the ``flowed Polyakov loop''. Further, 
we proceed to renormalize the Polyakov loop using 
this construct. Renormalization of Polyakov loop using Wilson flow has 
been considered earlier in Ref.~\cite{petreczky}, while a later paper
\cite{weber} has discussed various properties of the flowed Polyakov loop and 
renormalization of Polyakov loop. While we do not have place here to 
discuss those works, our approach to renormalized Polyakov 
is different from what has been followed there.
In the last section, we discuss the flow-time behavior
of various parts of the gluon condensate. It is known that as one
crosses the deconfinement temperature \tc\, the gluon condensate 
starts to melt. We find that the electric and magnetic components of the
condensate have very different flow behaviors, and use them to explore the
temperature dependence of the different condensates.

\section{Polyakov loop}

The deconfinement transition is associated with the breaking of the
$Z_3$ center symmetry for SU(3) gauge theory. The Polyakov loop,
\beq 
L(T, a) \ = \ \frac{1}{3 V} \ \sum_{\bf x} \ {\rm Tr}
\prod_{x_4=1}^{N_t} U_4 ({\bf x}, x_4)
\eeq{polloop} 
transforms nontrivially under the $Z_3$ symmetry and acts as an order
parameter for the transition. Following standard arguments, in finite volume
system one monitors the transition by looking at
\beq
P(T, a) \ = \ \langle |L(T, a)| \rangle_{\scriptscriptstyle T}.
\eeq{ploop}

The bare Polyakov loop, as defined in \eq{ploop}, depends strongly on 
the lattice spacing $a$ \cite{polyakov}, \\
\beq
P(T, a) \ = \ e^{-f(g^2(a))/aT} \ P_{\rm ren}(T),
\eeq{polren}
approaching $0$ as $a \to 0$. 
On the other hand, Wilson flow can be used to define an order parameter 
that is only mildly
dependent on the lattice spacing $a$, and has a finite continuum limit: if
we flow to a physical scale $t$, and define a Polyakov loop, $P(T, t,
a)$ through \eq{polloop} with the links replaced by flowed links, then 
$P(T, t, a) = P(T, t) + \mathcal{O}(a^2/t)$. Since the  
Wilson flow preserves center symmetry, the flowed Polyakov loop 
$P(T, t, a)$ can be treated as an order parameter for the deconfinement
transition.

\begin{figure}
\centerline{\includegraphics[width=.5\textwidth]{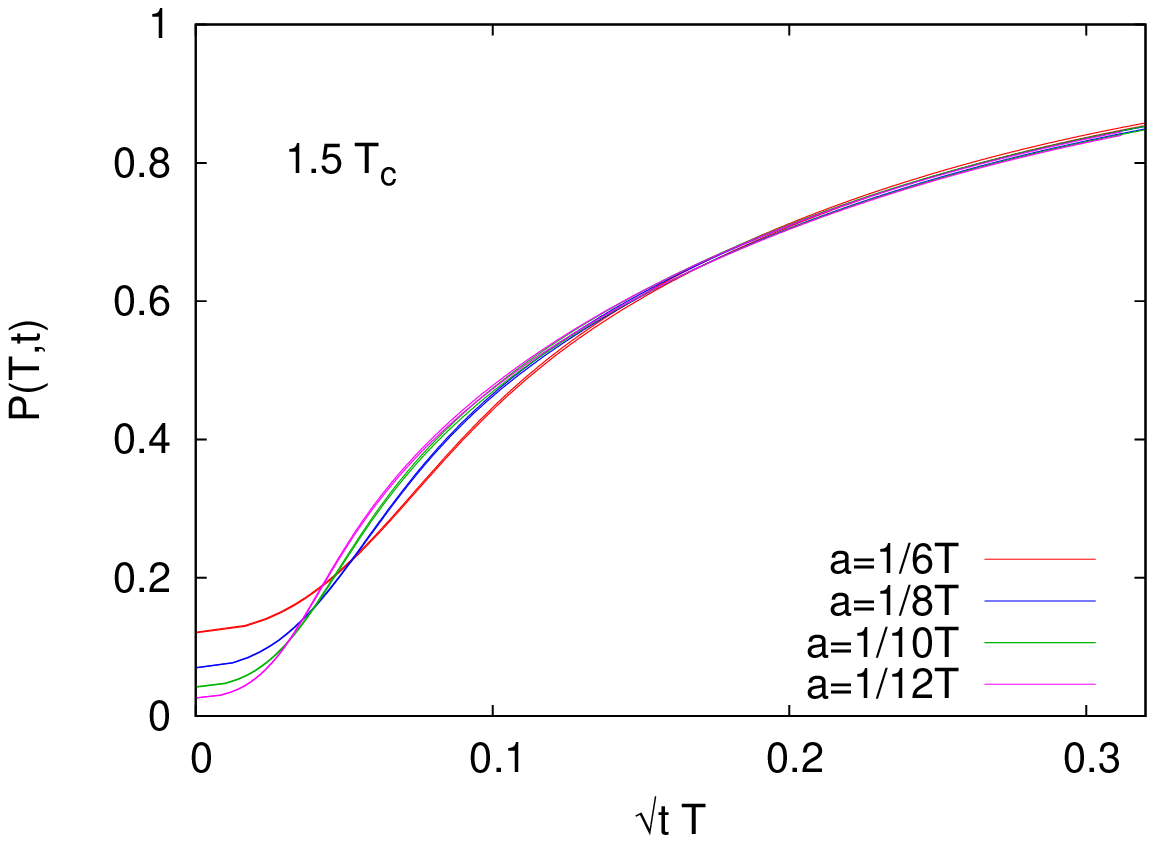}\includegraphics[width=.5\textwidth]{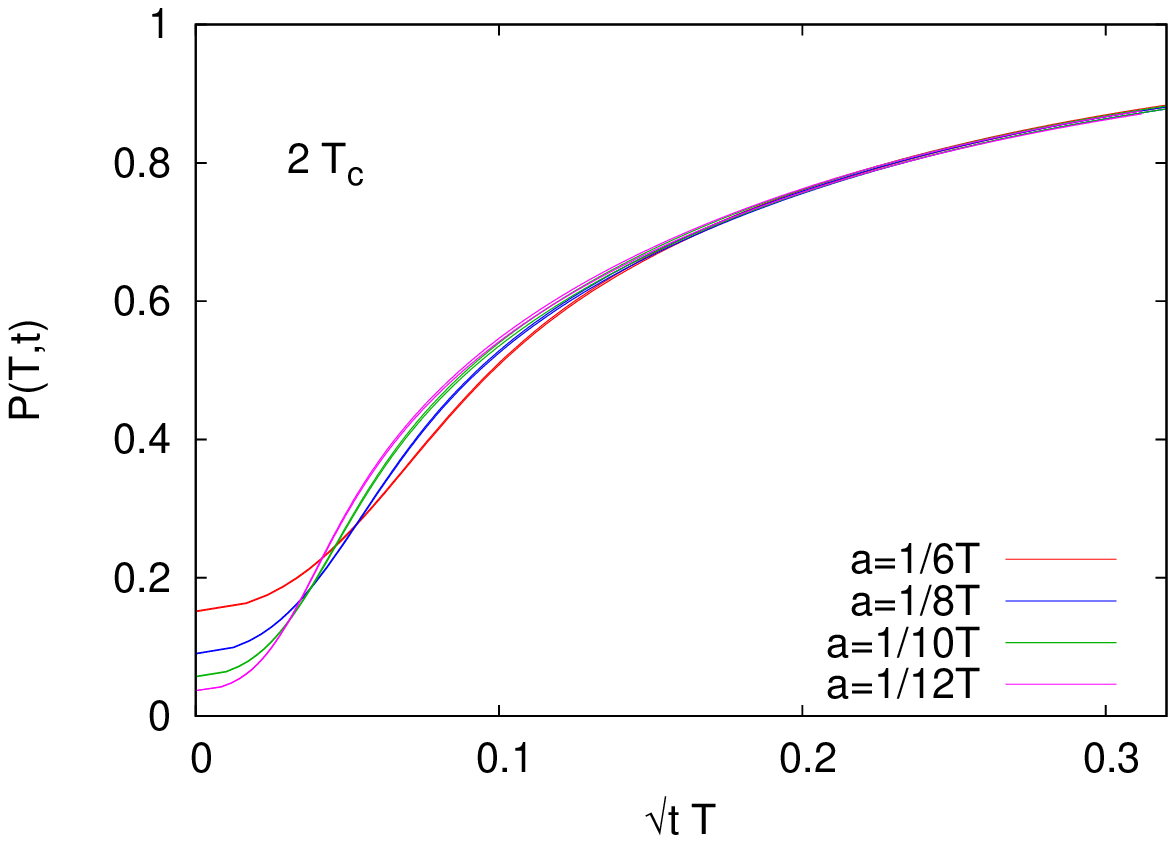}}
\caption{Flowed Polyakov loop at 1.5 \tc\ (left) and 2 \tc\ (right) as a function of flow time.} 
\label{fig1}
\end{figure}

In Fig.\ref{fig1}, we show the flowed Polyakov loop for four different
lattice spacings at two different temperatures. The strong $a$ dependence
at $t=0$, indicated by \eq{polren}, is removed at finite flow times;   
the remaining finite $a$ corrections are seen to be suppressed
when the flow time increases to $\sqrt{t} T \simeq 1/N_t$,
i.e, $\sqrt{t}/a \simeq 1$ on the respective lattices. This is less restrictive
than \eq{range}, and allows one to have a window where finite temperature
studies with Wilson flow can be performed on present day lattices.

\begin{figure}
\centerline{\includegraphics[width=.5\textwidth]{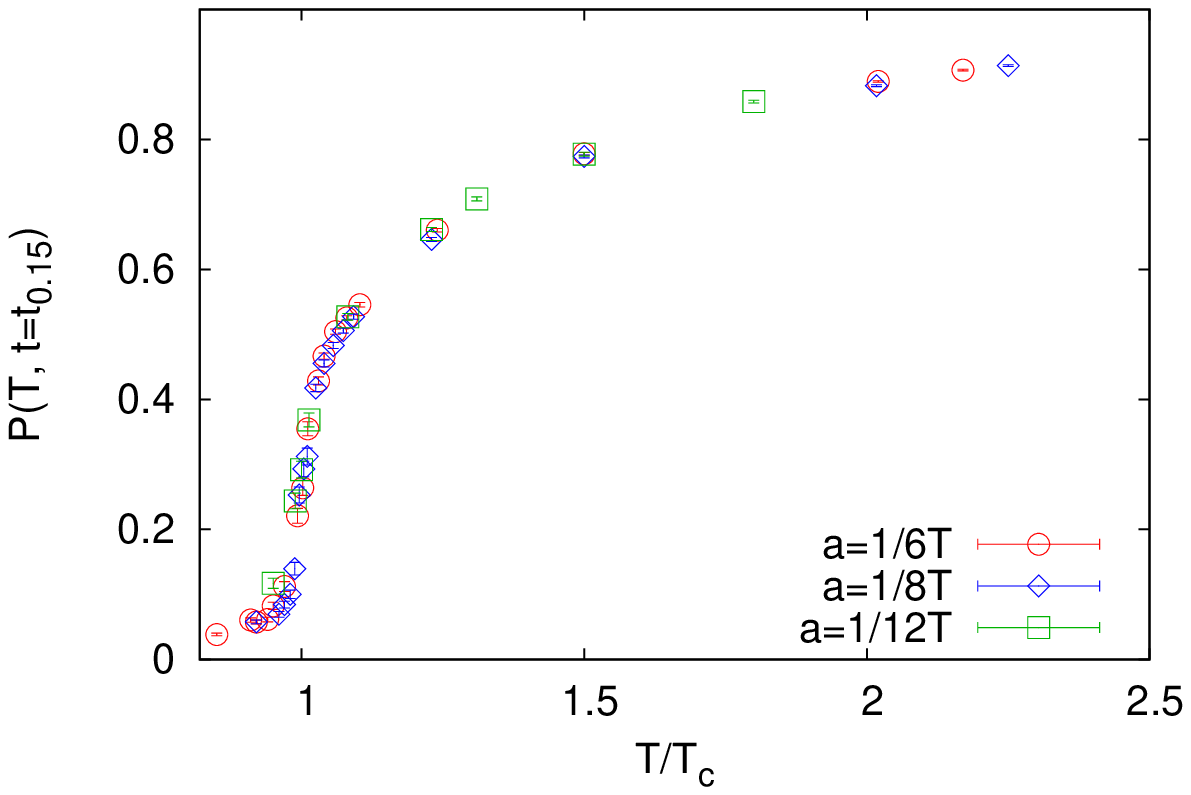}\includegraphics[width=.5\textwidth]{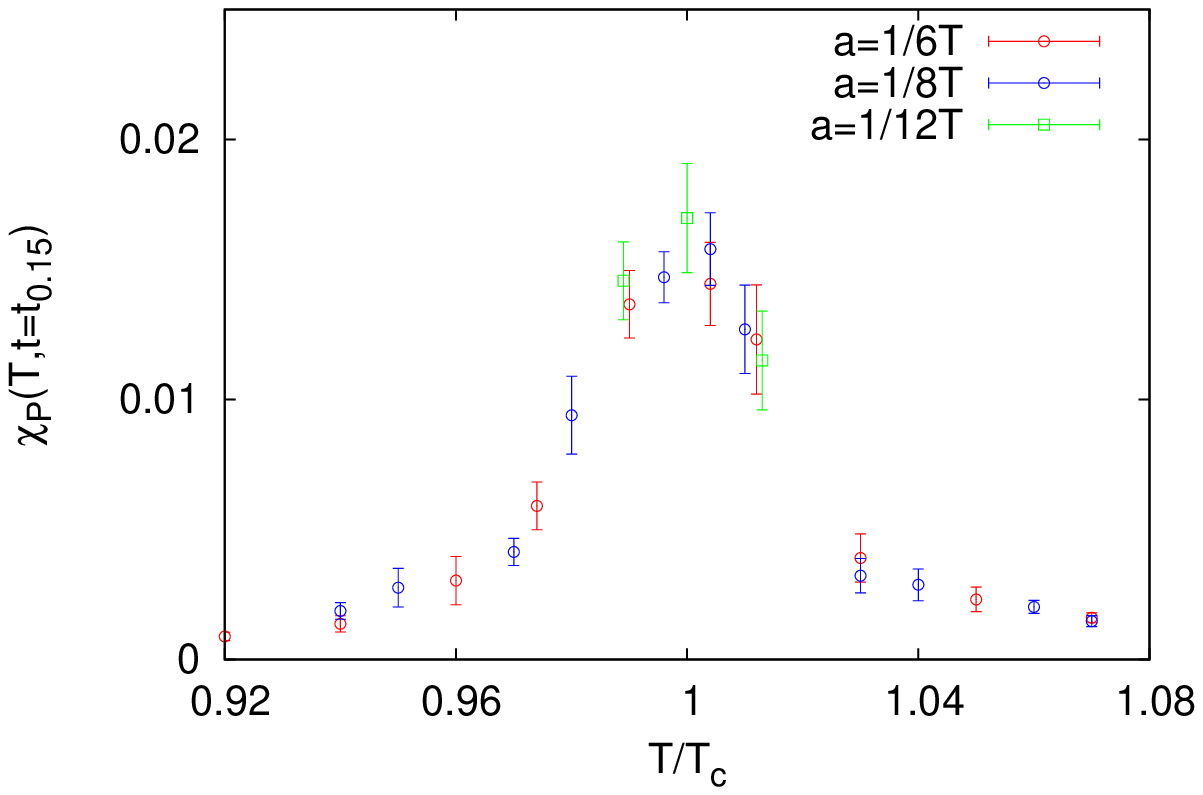}}
\caption{The flowed Polyakov loop at flow time $t_{0.15}$ (left) and its 
susceptibility density.}
\label{fig2}
\end{figure}

In Fig. \ref{fig2} the flowed Polyakov loop $P(T, t, a)$ is shown at a flow 
time $t=t_{0.15}$, at three different lattice spacings. This flow time is large
enough that the finite $a$ effects are negligible. From now on, we will 
restrict ourselves to flow times $t > 1/a^2$ such that finite $a$ effects
are negligible, and suppress the argument $a$, referring simply to $P(T, t)$. 
 
This $a$-independent flowed Polyakov loop is sufficient to measure the
continuum deconfinement transition in pure gauge theory. In the right 
panel of Fig. \ref{fig2} we show the susceptibility density \\
\beq
\chi_P (T, t) = \tav{|P(T, t)|^2} - \tav{|P(T, t)|}^2 .
\eeq{susc}
At a first order transition, $\chi_P (T, t)$,
is expected to show a volume-independent peak at $T_c$, just like the 
susceptibility density for the non-flowed loop. 
Unlike the latter, however, the flowed
susceptibility peak height does not change with $a$, as illustrated in the
figure.

We next turn to the extraction of the renormalized Polyakov loop $\pren$, 
\eq{polren}, from $P(T, t)$. Since with flow, $\sqrt{t}$ acts as an inverse 
momentum cutoff, analogous to \eq{polren} one can write 
\beq
\pren = \exp{\frac{R \left( g^2(t) \right) }{\sqrt{t} T}} \ P(T, t).
\eeq{prent}
where, to leading order, \cite{prd}
\beq
R \left(g^2(t) \right) = 
\frac{1}{3 \pi^2} \frac{\sqrt{\pi}}{\sqrt{8}} \ g^2(t) \ 
\left( 1 + \mathcal{O}(g^2) \right) .
\eeq{expr}
Following standard arguments \cite{polyakov} we expect that $\pren$ 
is a function of temperature up to $\mathcal{O}(\sqrt{t} T)$ corrections, 
and has a finite limit as $t \to 0$. 

\begin{figure}
\centerline{\includegraphics[width=.5\textwidth]{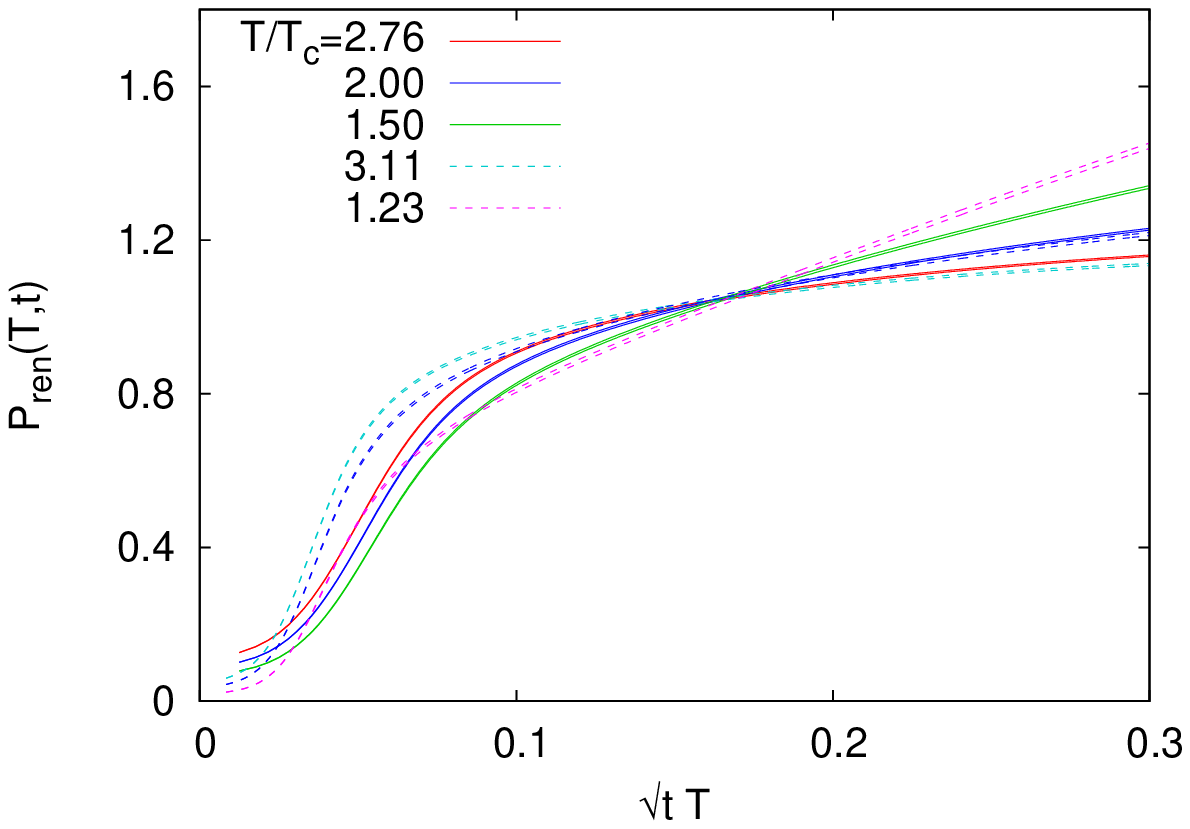}
\includegraphics[width=.5\textwidth]{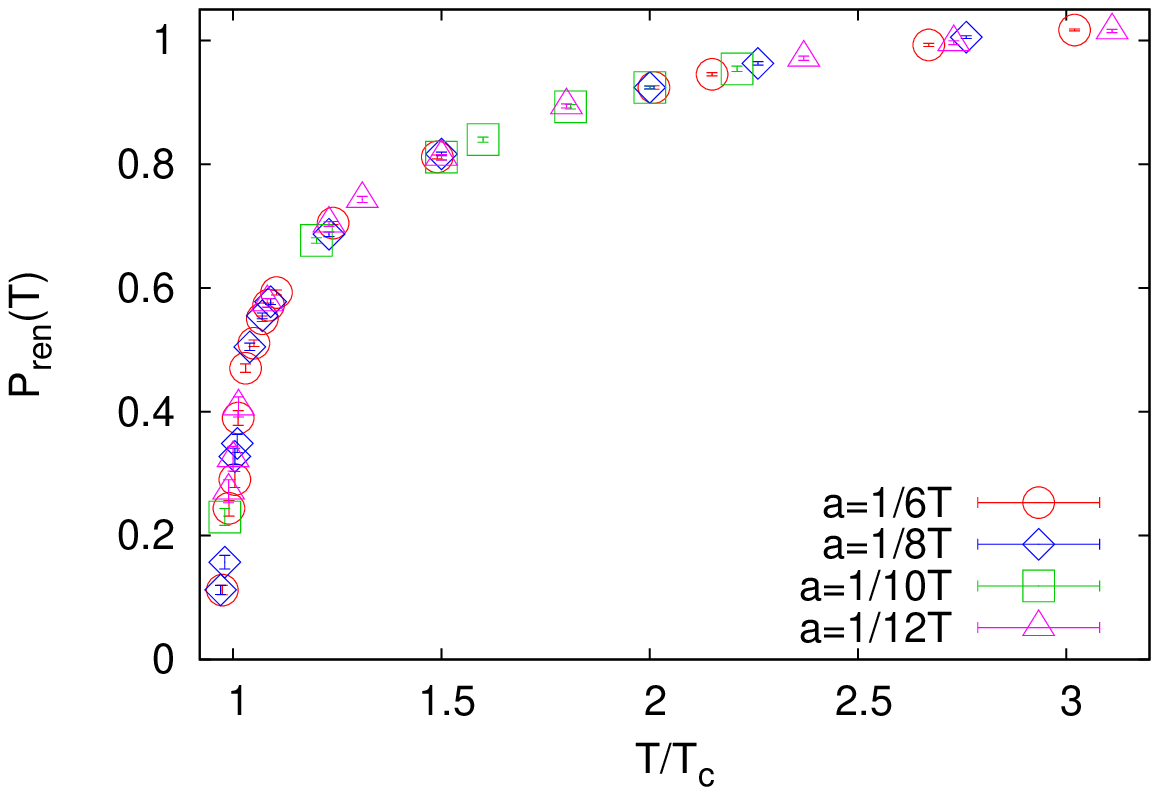}}
\caption{(Left) $\prenT$, calculated using Eq.(2.6), at a few 
temperatures. The solid lines show calculations on 
lattices with $a=1/8T$ and the dashed lines, those at $a=1/12T$.
(Right) Renormalized Polyakov loop from a nonperturbative matching. 
See text.}
\label{fig3}
\end{figure}

In order to calculate $\pren$ using \eq{expr}, we need to evaluate
$g^2$.  In this report we will evaluate $\gms$ by inverting
\eq{coupling}. See Ref.  \cite{prd} for results with a direct
evaluation of $\gms$ from $\Lambda_{\overline{\scriptscriptstyle
    MS}}$, and comparison with $\gms$ from \eq{coupling}. In the left
panel of Fig. \ref{fig3}, we show calculations of $\pren$ using \eq{expr}
at different temperatures, for lattices with $a=1/8T$ and $1/12T$,
respectively. While at higher temperatures, the remnant flow time
dependence is mild, we see that as one comes closer to \tc\, it
becomes much stronger and an extrapolation to $t \to 0$ becomes
difficult. Also while the finer lattice shows clear
improvement at higher temperatures, it still is not good enough to extract
$\pren$ close to $T_c$.

In order to calculate $\pren$ at lower temperatures, we therefore 
follow a different strategy. Note that the temperature dependence of the
renormalization factor is simple, \eq{prent}, and therefore, the
renormalization factor at a lower temperature can be simply obtained from
the renormalization factor at a higher temperature up to remnant
linear $\sqrt{t} T$ corrections, which we expect to be small if we
remain within a window $\sqrt{t} T \in (0.2, 0.3)$. In order to
extract $R \left( g^2(t) \right)$, we take, as a baseline value,
$P_{\rm ren}(3 T_c)$ = 1.0169(1) \cite{renpol1}.
This determines $R \left( g^2(\sqrt{t}=1/10 T_c) \right)$,
which can then be used to calculate $P_{\rm ren}$ to all temperatures
upto 2 \tc\. This process is then iterated to calculate $P_{\rm ren}$
at lower temperatures.  This strategy is similar in spirit to that
followed in Ref.~\cite{renpol1}; however, the use of flow makes the
calculation simpler, as we do not need to match lattices at different
lattice spacings to same temperature. The renormalized Polyakov loop
extracted this way is shown in Figure \ref{fig3}.

\section{Gluon condensates}

The nonperturbative nature of the QCD vacuum is characterized by various 
condensates. The dimension four, scalar condensate ${\bf E}$ can give rise
to two operators at finite temperature,
 \beq
E = {\rm Tr} G_{0i} G_{0i}, \qquad M = \frac{1}{2} {\rm Tr} G_{ij} G_{ij}
\eeq{ftcondop}
connected by O(4) transformations.

\begin{figure}
\centerline{\includegraphics[width=.5\textwidth]{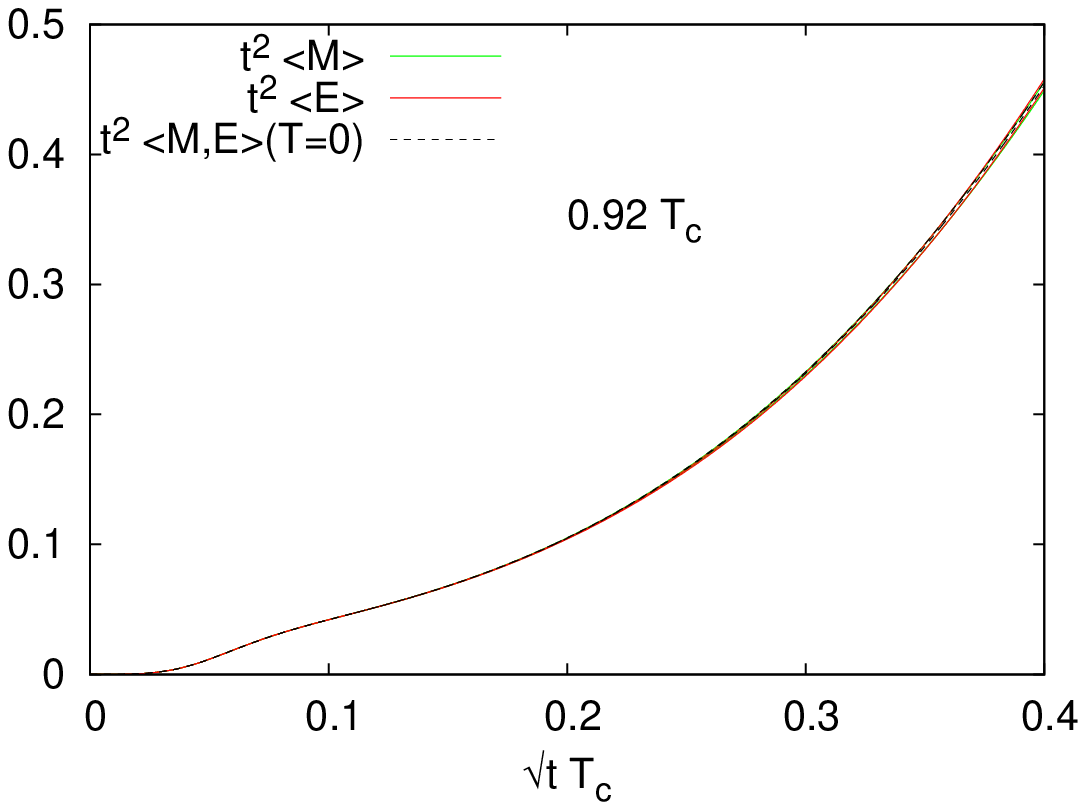}\includegraphics[width=.5\textwidth]{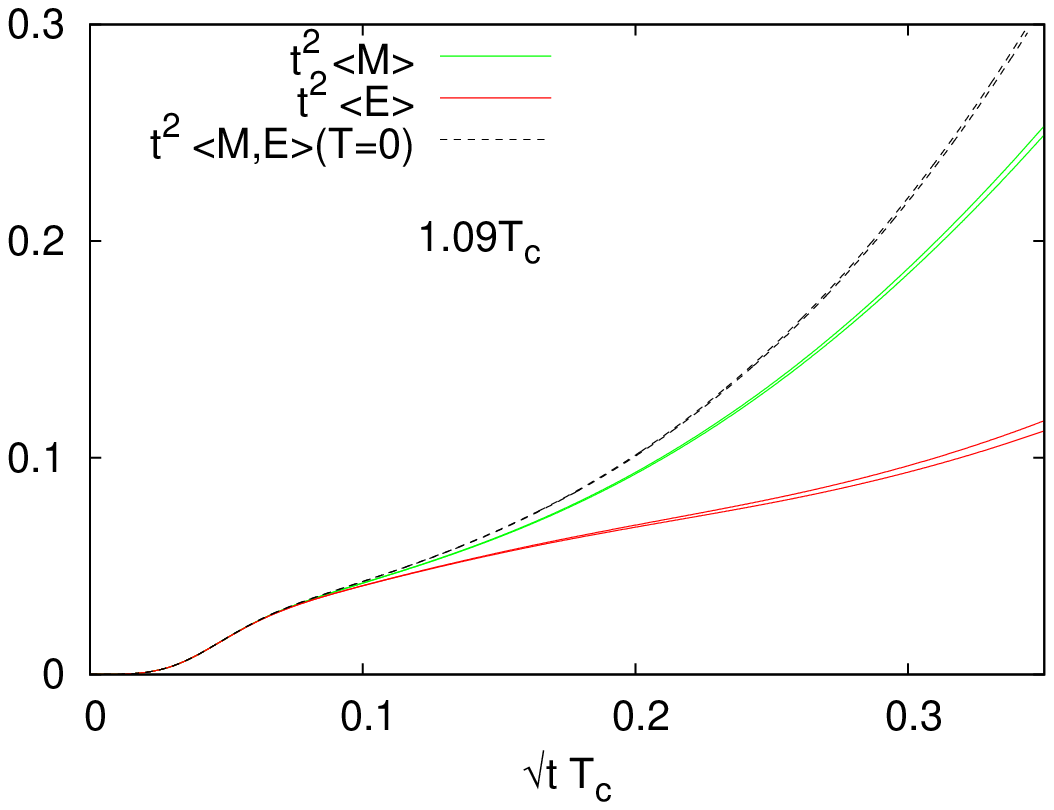}}
\caption{The electric and magnetic condensate operators $t^2 \langle
  M, E \rangle$ plotted against flow time, at temperatures of 0.92
  $T_c$ (left) and 1.09 $T_c$ (right). Also plotted are the zero 
temperature values of the same operators.}
\label{fig5}
\end{figure}

The flow behaviors of $E$ and $M$ turn out to be quite interesting. 
In Fig. \ref{fig5} we show the dimensionless flowed quantities 
$t^2 \Eav$ and $t^2 \Mav$ immediately below and above $T_c$, together
with the corresponding operator at T=0. The figure indicates that O(4)
symmetry breaking sets in rather abruptly in a narrow temperature interval
near $T_c$.

\begin{figure}
\centerline{\includegraphics[width=.5\textwidth]{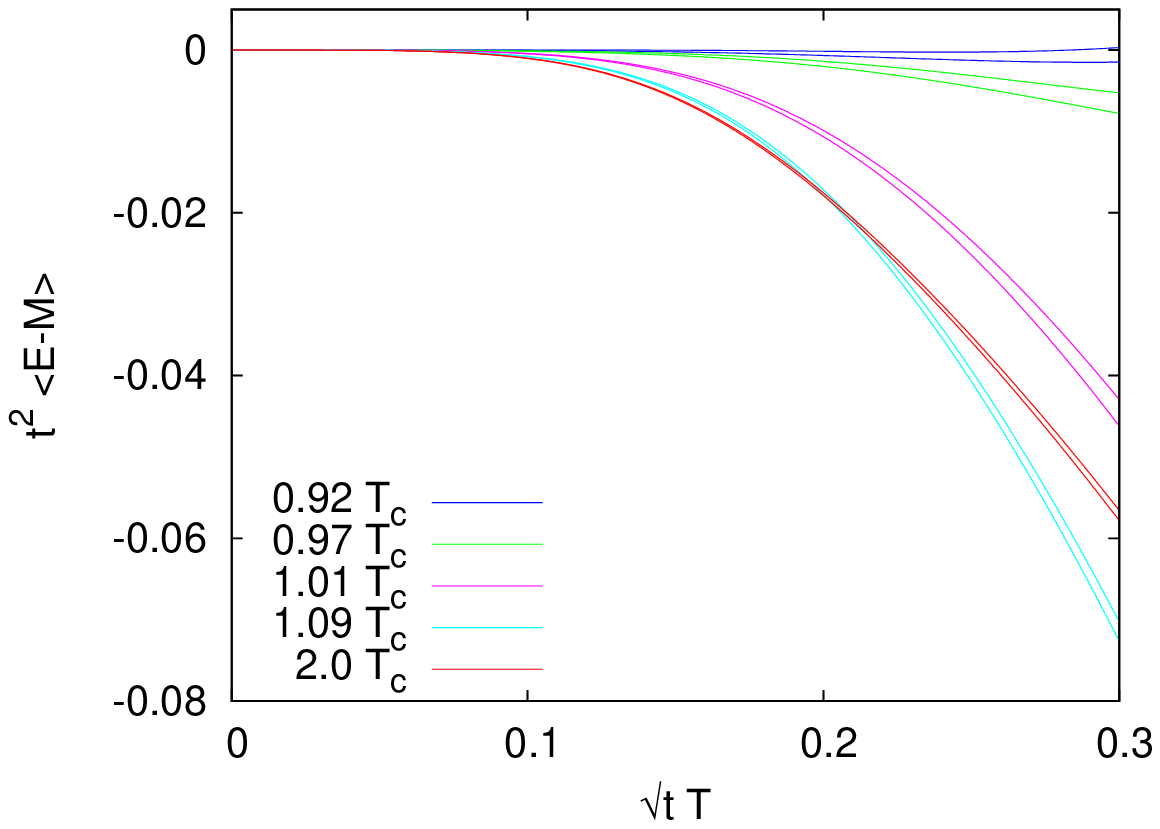}\includegraphics[width=.5\textwidth]{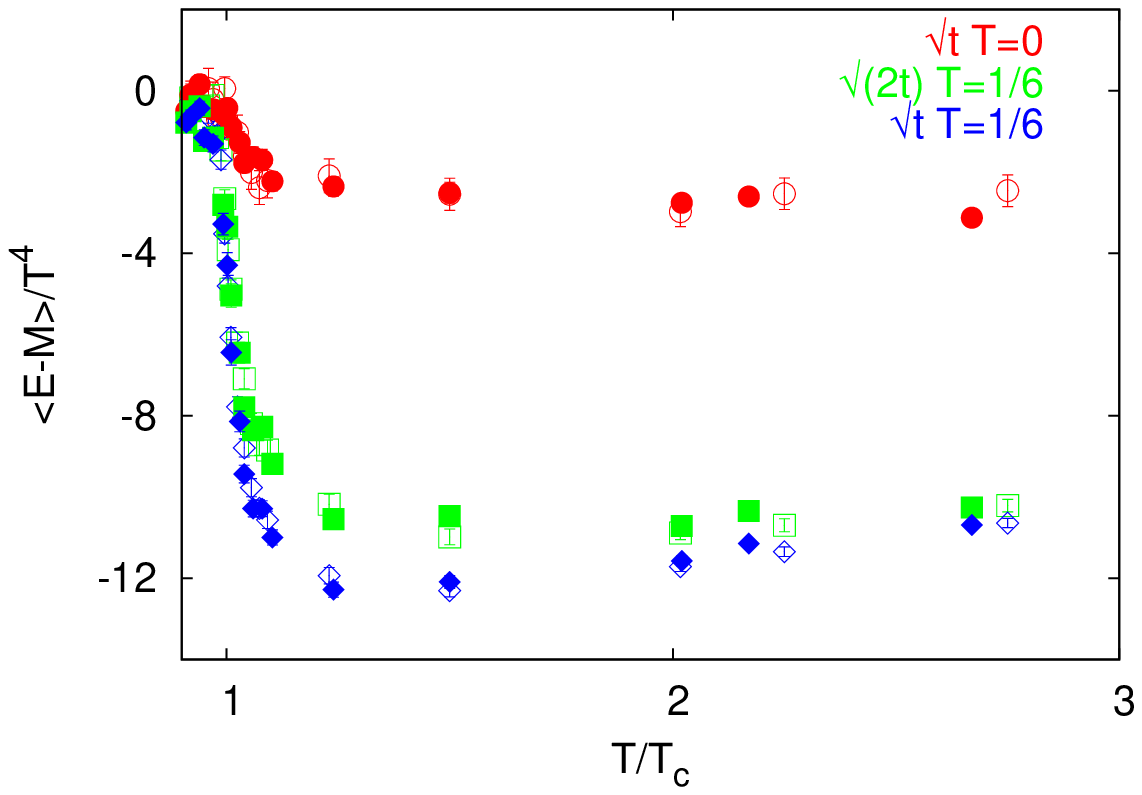}}
\caption{(Left) Flow time dependence of the operator $t^2 \langle E -
  M \rangle$ at different temperatures, for lattices with
  $N_t=8$. (Right) Temperature dependence of the condensate difference
  $\langle E(T, t) - M(T, t) \rangle /T^4$, at zero flow and at two
nonzero flow times, for $N_t=8$ (empty symbols) and $N_t=6$ (filled) 
lattices.}
\label{fig6}
\end{figure}

In the left panel of Fig. \ref{fig6} we show the difference 
$t^2 \langle E(T,t) - M(T,t) \rangle$ . At large flow times
$\tT \gtrsim 1/N_t$ this quantity is very
sensitive to the deconfinement transition, remaining very small
upto \tc\ and then showing a jump.  
In the right panel of the figure we show $\langle E(T, t) - M(T, t) 
\rangle /T^4$, for $t=0$ and two different nonzero flow times. 
The flow behavior of $\Eav$ and $\Mav$ lead to a sharp jump in this object,
which can be used to monitor the traaansition.

\begin{figure}
\centerline{\includegraphics[width=.5\textwidth]{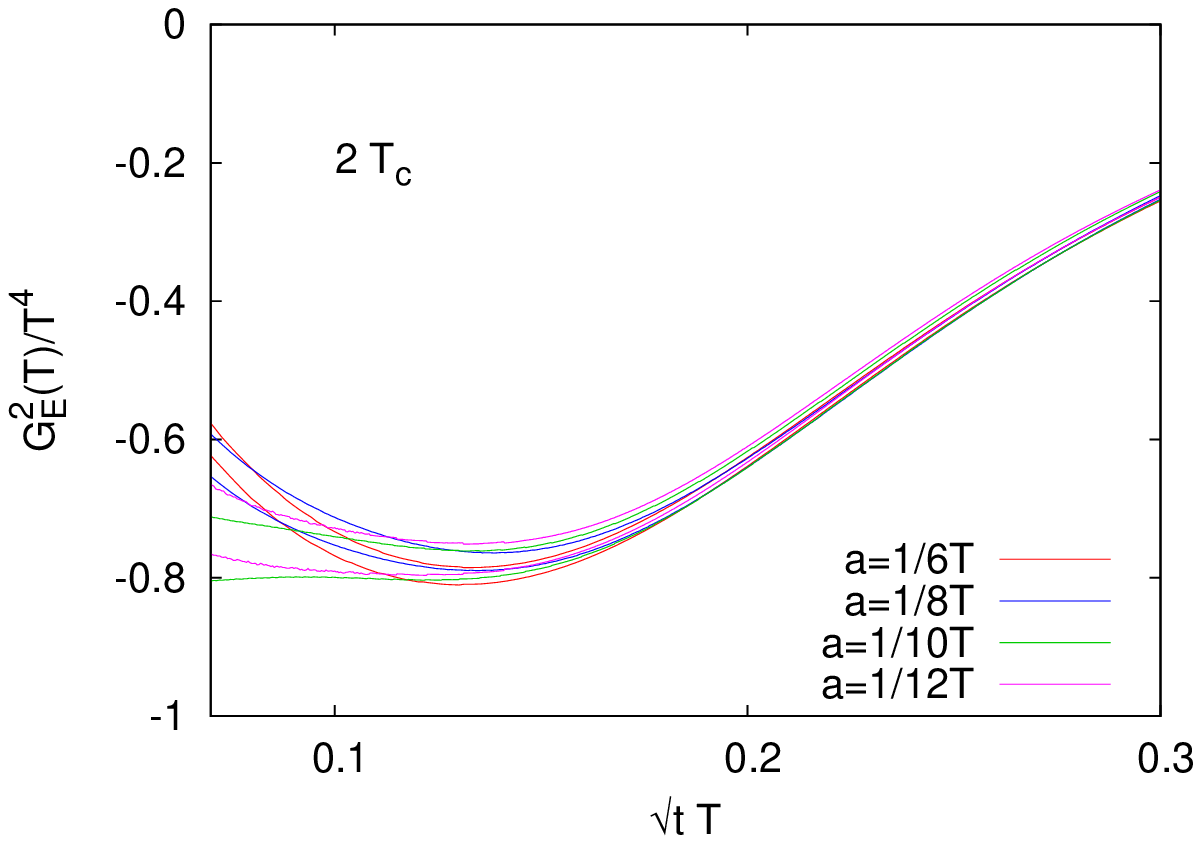}\includegraphics[width=.5\textwidth]{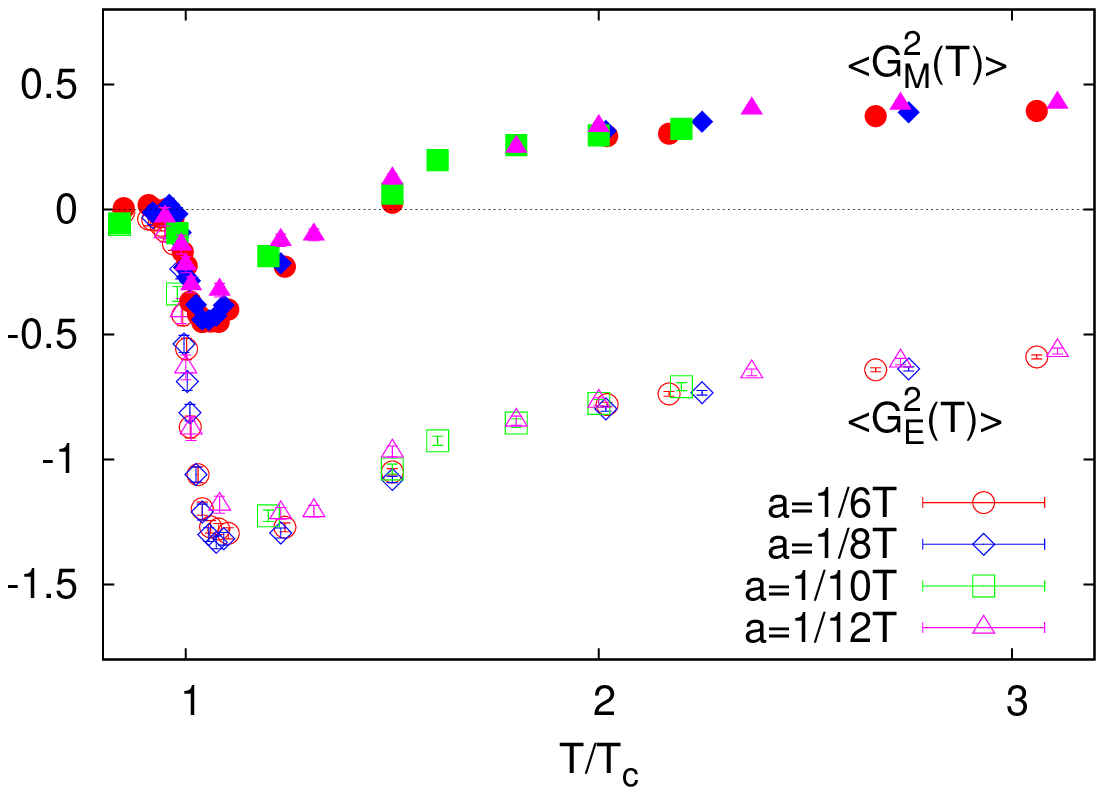}}
\caption{(Left) Flow time dependence of the electric gluon condensate, Eq.(3.2). (Right) Estimate of the renormalized, vacuum subtracted electric and magnetic gluon condensates. at different temperatures.}
\label{fig7}
\end{figure}

The flowed condensates can be used to calculate the conventionally
normalized gluon condensates. For this, we first do a vacuum subtraction, i.e.,
calculate $\underline{E}(T)=E(T)-E(0)$ and similarly for $M$. Then the 
vacuum subtracted finite temperature gluon condensates can be written as
\cite{suzuki}
\beq
\langle \ \underline{G^2_{\scriptscriptstyle E,M}}(T) \ \rangle = \lim_{t \to 0} 
\frac{1}{\pi^2} \ R(t) \ \tav{\underline{E}, \underline{M}}(T, t), \qquad R(t) \ = \ 1 \, - \, 0.1116 \, \frac{11}{16 \pi^2} \ 
\gms(\mu=\frac{1}{\sqrt{8 t}}) \ + \ \mathcal{O} (g^4) 
\eeq{gcond}
In the left panel of 
Fig. \ref{fig7} we show, for illustration, $\Ebar (2 T_c)$ at different flow 
times. In the coarser lattices, the result is rather disappointing: there is 
hardly any hint of a plateau or a linear behavior from which one can extract
the $t \to 0$ limit. On the other hand, with the two finest lattices, a plateau 
starts to form. Further, in the two coarser lattices, the condensate is in 
agreement with the finer lattices up to this plateau region, but then the 
lattice spacing effects set in as $\sqrt{t}$ becomes $\lesssim a$ before 
formation of a proper plateau. 
Taking the value in this region to be an approximation to the 
$t \to 0$ limit, we show in the right panel of the figure the value of the
vacuum subtracted condensates. The result is quite interesting: just above 
\tc\ both the condensates show large values. At higher temperatures, while 
the electric and magnetic condensates themselves do not become insignificant,
they have opposite signs, which lead to a small $G^2$.

The computations reported here were carried out in the gaggle and pride 
clusters of the department of theoretical physics, TIFR. We thank Ajay Salve 
and Kapil Ghadiali for technical support.


\begin{thebibliography}{99}

\bibitem{main}
 M. L\"uscher, \jhep 1008 (2010) 071.
\bibitem{narayanan}
 R.\ Narayanan and H.\ Neuberger, \jhep 0603 (2006) 064.
\bibitem{scale}
  See, e.g., R. Sommer, PoS LATTICE {\bf 2013} (2014) 015 (arXiv:1401.3270), 
and references therein.
\bibitem{luscher}
M. Luscher, \jhep 1304 (2013) 123.
\bibitem{suzuki}
H. Suzuki, \ptep (2013) 083B03; \ptep (2015) 103B03.
\bibitem{flowqcd}
  M. Asakawa, T. Hatsuda, E. Itou, M. Kitazawa and H. Suzuki, 
  \pr D90 (2014) 011501; \pr D 92 (2015) 059902.
\bibitem{prd}
  S. Datta, S. Gupta and A. Lytle, arXiv:1512.04892, to be published in
\pr D.
\bibitem{petreczky} 
  P. Petreczky and H.-P. Schadler, \pr D 92 (2015) 094517.
\bibitem{weber}
  A. Bazavov, et al., \pr D 93 (2016) 114502.
\bibitem{polyakov} A.M. Polyakov, \np B164 (1980) 171.
\bibitem{renpol1}
  S. Gupta, K. H\"ubner and O. Kaczmarek, \pr D 77 (2008) 034503.
\end{thebibliography}
\end{document}